\documentclass{emulateapj}
\usepackage{graphicx,natbib}

\begin{document}

\title{Observational signatures of simulated reconnection events in the
solar chromosphere and transition region}

\author{L. Heggland}
\affil{Institute of Theoretical Astrophysics, University of Oslo,
  P.O. Box 1029, Blindern, N-0315 Oslo, Norway}
\email{lars.heggland@astro.uio.no}

\author{B. De Pontieu}
\affil{Lockheed Martin Solar and Astrophysics Laboratory, 3251 Hanover Street,
  Org. ADBS, Building 252, Palo Alto, CA 94304}

\author{V. H. Hansteen\altaffilmark{1}}
\affil{Institute of Theoretical Astrophysics, University of Oslo,
  P.O. Box 1029, Blindern, N-0315 Oslo, Norway}

\altaffiltext{1}{Also at Center of Mathematics for Applications,
  University of Oslo, P.O. Box 1053, Blindern, N-0316 Oslo, Norway}

\begin{abstract}

We present the results of numerical simulations of wave-induced
magnetic reconnection in a model of the solar atmosphere. In the magnetic field
geometry we study in this article, the waves, driven by a monochromatic
piston and a driver taken from Hinode observations, induce periodic
reconnection of the magnetic field, and this reconnection appears
to help drive long-period chromospheric jets. By synthesizing
observations for a variety of wavelengths that are sensitive to a wide
range of temperatures, we shed light on the often confusing
relationship between the plethora of jet-like phenomena in the solar
atmosphere, e.g., explosive events, spicules, blinkers, and other
phenomena thought to be caused by reconnection.

\end{abstract}


\section{Introduction}

The Sun displays a bewildering array of jet-like phenomena that can be
observed at many different wavelengths and in many different regions.
At the limb, we find
spicules, protrusions of cool gas that can reach heights of
6-10~Mm and have lifetimes of several minutes \citep[e.g.][]{Beckers1968},
and that are observed in chromospheric lines such as H$\alpha$ and
Ca~{\sc ii}~H. There are also larger and longer-lasting jets such as surges
and macrospicules. Recently, faster
and shorter-lived jets, called type II spicules, have been found
\citep{DePontieu+etal2007b}, many lasting less than 100~s before fading
from view. Hinode observations were instrumental in finding these, but
have also shown a large amount of other jet activity in Ca~H and other lines
\citep[e.g.][]{Katsukawa+etal2007,Shibata+etal2007}.

On the disk,
mottles show much of the same behaviour as spicules, and we can also observe
shorter-lived dynamic fibrils in the same lines. In higher-temperature
ultraviolet lines such as C~{\sc iv}, Si~{\sc iv}, and O~{\sc vi},
we observe short-lived explosive events and
longer-lasting blinkers. The former, in particular, have very wide line
profiles indicating strong bidirectional jets \citep{Innes+etal1997}.

There have been investigations into whether some of these terms simply
describe different aspects of the same basic phenomena, e.g. spicules and
mottles \citep{Grossmann-Doerth+Schmidt1992,Tsiropoula+etal1994}, blinkers
and explosive events \citep{Chae+etal2000}, spicules and explosive events
\citep{Wilhelm2000}, or spicules, blinkers, and explosive
events \citep{Madjarska+Doyle2003}. Different authors often conclude
differently, and the matter can not be considered settled. Part of the
problem is that we usually lack cotemporal, cospatial observations across
several wavelength bands, so that we most often cannot trace events across
different temperatures and with enough spatial and temporal resolution to
elucidate the relationship between the various features. Another complication
is that we cannot observe the exact same phenomenon both on the disk
and at the limb, for simple geometric reasons. Simulations can be very helpful
here in letting us change our vantage point at will.

The driving mechanism of the various jets is in general not well established.
The main candidates have been waves (magnetoacoustic or Alfv\'{e}n) and
magnetic reconnection. There are strong indications that acoustic shockwaves
cause dynamic fibrils \citep{Hansteen+etal2006,DePontieu+etal2007,
Heggland+etal2007} and that reconnection causes explosive events
\citep{Innes+etal1997,Chae+etal1998a}, but the case is more open for
the other types of jets. It has been noted by e.g. \citet{Chae+etal1998a},
\citet{Ning+etal2004}, and \citet{Doyle+etal2006} that many explosive events
occur in bursts, often with intervals of 3-5~minutes. These periods correspond
to the dominant wave modes produced by the solar granulation, suggesting
that such waves may induce or modulate reconnection events.

A review of various spicule models can be found in \citet{Sterling2000}.
Most models so far are in 1D, assuming a rigid magnetic flux tube either with
or without expansion. Some use a piston driver, either at a given frequency
or randomised, while others
use a sudden increase in pressure and temperature as a trigger mechanism
for setting up shocks. This pressure increase is often assumed to be in
the photosphere, but \citet{Shibata+etal1982} and \citet{Sterling+etal1993}
have investigated the effects of energy input in the upper chromosphere as
well, producing different types of jets that have similar properties to
spicules or surges. The source of the energy input is usually not
specified, though reconnection could be a natural candidate, as speculated
by \citet{Sterling+etal1993}.

2D simulations containing an actual magnetic field (as opposed to treating
it merely as a given rigid flux tube) have been performed by
\citet{Takeuchi+Shibata2001a,Takeuchi+Shibata2001b}. They studied
photospheric reconnection which they claim
produces a large enough wave energy flux to drive spicules, but their
actual simulation box does not extend past the lower chromosphere.

Explosive events have been modelled in 1D by \citet{Erdelyi+etal1999}
and \citet{Sarro+etal1999}. 2D simulations of explosive events have been
performed by \citet{Karpen+etal1995}, \citet{Innes+Toth1999},
\citet{Roussev+etal2001a,Roussev+etal2001b,Roussev+etal2001c},
and \citet{Chen+Priest2006}. Again, the 1D simulations
generally assume a rigid flux tube geometry, and treat the magnetic
reconnection not explicitly, but as a sudden deposition of energy at a
specified height. The 2D simulations often assume a simplified magnetic
geometry with
vertical antiparallel field lines. \citet{Karpen+etal1995} use a more
complicated magnetic field geometry with some similarities to the one
used in this paper, but do not include radiative losses or heat conduction
in their simulations.

In this paper, we present the results of 2D simulations that include the
effects of radiation and heat
conduction, and that involve a complex magnetic field geometry in which
reconnection events are induced by waves propagating upwards from
the photosphere/convection zone. We do not attempt to match the results
to specific observations, but show the signatures of these events in
several chromospheric, transition region and lower coronal spectral lines,
and viewed both on the disk and at the limb of the Sun.

\section{Simulations}

\begin{figure}
 \begin{center}
  \plotone{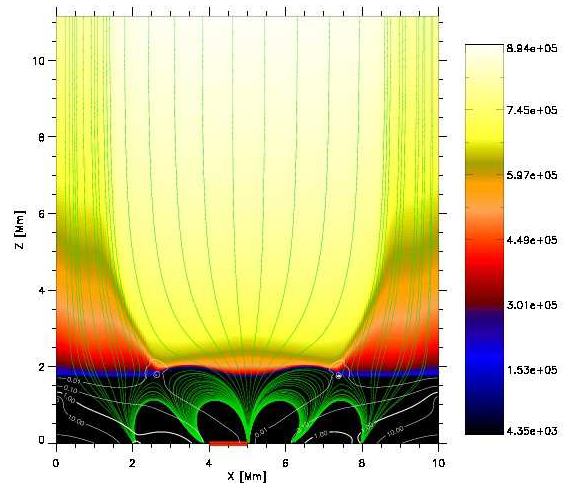}
 \end{center}
 \caption{Initial temperature structure, with overplotted $\beta$ contours
          ({\it white}) and magnetic field lines ({\it green}). The red line
          between
          $x=4$ and $x=5$~Mm marks the location of the piston. The magnetic
          null points show up as local maxima of the $\beta$ at $z = 1.75$~Mm
          and $x = \{2.6,7.4\}$~Mm.}
\label{initatm}
\end{figure}

The model atmosphere is similar to the one used by \citet{Heggland+etal2007},
but now the lower boundary of the domain reaches down to the upper photosphere,
as opposed to the chromosphere. The simulation box extends about 11~Mm in
height, going from the photosphere through the chromosphere, transition region
and lower corona. The lower boundary, $z=0$, corresponds to a height
of 150~km in the VAL3C model \citep{Vernazza+etal1981}. The simulation box
contains $201 \times 191$ grid cells, using a uniform spacing of 50~km in the
horizontal direction and a non-uniform spacing in the vertical direction,
starting at 16~km in the chromosphere and transition region and increasing
exponentially in the corona, reaching 220~km at the upper boundary.
In Figure~\ref{initatm}, we have plotted the initial temperature
distribution with magnetic field lines and curves of equal plasma $\beta$,
the latter defined as the ratio between the thermal and the magnetic pressure.

\begin{figure}
 \begin{center}
  \plotone{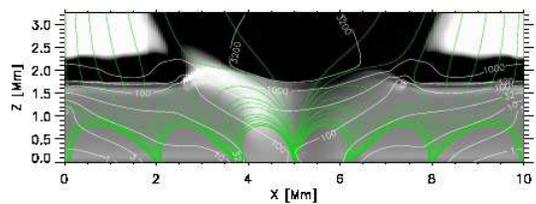}
 \end{center}
 \caption{Plot of the vertical velocity, set to saturate at 1~km~s$^{-1}$.
          Selected magnetic field lines in green, Alfv\'{e}n speed
          contours in km~s$^{-1}$ in white.
          The first fast mode wave reaches the transition region after 12
          seconds, refracts away from the high Alfv\'{e}n speed in
          the centre, and perturbs the field in the vicinity of the
          magnetic null points, triggering reconnection.}
 \label{firstwave}
\end{figure}

The magnetic field configuration is a potential field, with
a central flux tube which progressively widens with height and
ends up dominating the field in the corona. However, there are also two
inclined flux tubes, one on each side of the centre. These tubes both
end up in magnetic null points, located at transition region heights
($z = 1.75$~Mm, $x = 2.6$~Mm and $x=7.4$~Mm). The null points can
be seen in Figure~\ref{initatm} as $\beta$ maxima that are very
small in spatial extent.
It is near these null points that the reconnection events we study in
this paper occur. Finally, there are two outer flux tubes that open
into the corona.

It should be noted that this null point configuration is somewhat peculiar
to the 2D geometry, and would be difficult to reproduce exactly in a 3D
model, or indeed on the real Sun. However, it is our belief that the
observational
signatures of reconnection events on the real Sun should not be strongly
dependent on the specific geometry of the magnetic field, and that our
simulated observations should be broadly similar to what we would expect
to see when observing actual reconnection events. This is supported by
the similarities we find with observations of several types of events thought
to be caused by reconnection.

The code we use is the same 3D MHD code that was used in
\citet{Heggland+etal2007}, and is described in more detail by
\citet{Hansteen2005}. It should be noted that it contains a realistic
radiative loss function based on collisional excitation of hydrogen, carbon,
oxygen, neon, and iron, as well as thermal conduction along the magnetic
field (see \citet{Heggland+etal2007} for the details).

In the models presented here we operate with a magnetic diffusivity $\eta$
many orders of magnitude larger than that on the Sun, and dissipation starts at
much smaller magnetic field gradients. The dissipated energy is
\begin{equation}
Q_{\rm Joule}={\bf E}\cdot{\bf J},
\end{equation}
where ${\bf J}=\nabla\times{\bf B}$ is the current density and the resistive
part of the electric field is given by
\begin{equation}
E_x^{\eta}=\left\{{1\over 2}(\eta^{(1)}_y+\eta^{(1)}_z)
    +{1\over 2}(\eta^{(2)}_y+\eta^{(2)}_z)\right\}J_x,
\end{equation}
and similar for $E_y$ and $E_z$. The diffusivities are given by
\begin{eqnarray}
\eta^{(1)}_j&=&{\Delta x_j\over{\rm Pr}_M}(v_1c_f+v_2 |u_j| ) \\
\eta^{(2)}_j&=&{\Delta x_j^2\over{\rm Pr}_M}v_3|\nabla_\perp\cdot{\bf
u}|_{-},
\end{eqnarray}
where ${\rm Pr}_M$ is the magnetic Prandtl number, $c_f$ is the fast mode
speed, $v_1$, $v_2$ and $v_3$ are dimensionless numbers of order unity and the
other symbols retain their usual meanings.

The upper coronal boundary is maintained at 1 MK, while radiation and
heat conduction set the temperature structure in the rest of the domain.

In the simulations, we investigate the effects of different locations
and methods of driving, and how the produced waves can trigger reconnection.
In one case, we
use a monochromatic localised piston with 300 s period and 1.1~km s$^{-1}$
amplitude, driving the lower boundary in the vertical direction
at $z = 0$~Mm between $x = 4$~Mm and 
$x = 5$~Mm (marked with a red line in Fig.~\ref{initatm}). In another, we use a
driver taken from observations with Hinode/SP (Lites, private communication), 
producing waves at all locations along the lower boundary and in a more
realistic range of periods. The data were obtained on 2007 October 22
and contain the line-of-sight velocities in a slice across a network element. 

We also experiment with putting the null points at different heights
in the atmosphere. In one
case, they are located in the transition region at a height of about
1.75~Mm above the lower boundary, as in Figure~\ref{initatm};
in another, they are in the upper chromosphere, some 450~km lower.

\section{Analysis}

\subsection{Transition region null points, piston driver}
\label{hpcase}

\subsubsection{General description of events}

\begin{figure*}[tbp]
 \begin{center}
  \plotone{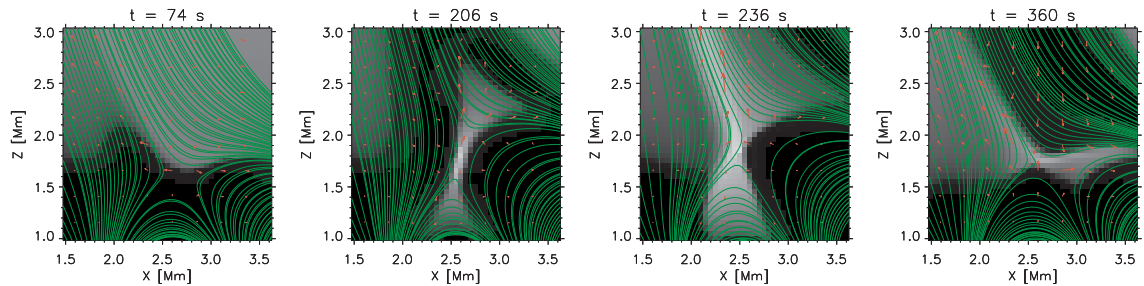}
 \end{center}
 \caption{Plots of the plasma temperature ({\it greyscale}) with magnetic field
          lines ({\it green}) and velocity vectors ({\it red}) at four
          different times
          in the model with transition region null points and piston driver.
          The plots zoom in on the area around the left-hand null point.
          At 74~s ({\it left panel}), a horizontal bidirectional jet is pushing
          cool material upwards to the left. At 206~s, reconnection has
          heated a significant area and propels a jet mainly across the
          field lines. At 236~s, the jet aligns more with the field and
          hot material is ejected upwards. The final panel shows a later
          horizontally oriented reconnection after 360~s. The field lines
          in all panels are drawn to show the geometry rather than the
          strength of the magnetic field.}
 \label{rjsecs}
\end{figure*}

We first look at the case with null points in the transition region and
the localised piston driver as described above. Due to the varying magnetic
inclination in
the region of the piston, it will generate a mixture of fast and slow mode
waves. In low-$\beta$ regions, which include the piston location as well as
most of the simulation box,
the fast modes can propagate everywhere, though they tend to be refracted
into regions of low Alfv\'{e}n velocity \citep[e.g.][]{Osterbrock1961},
such as null points.
The slow modes are mainly restricted to propagation along the magnetic field.

The waves disturb the field as they travel, and once the disturbance
reaches the null points, reconnection is triggered as lines of opposite
polarity are pushed together and meet. This process releases a significant
amount of energy in
a short time and leads to very rapid heating of the plasma; in the stronger
events, the plasma around the null points reaches coronal temperatures of
roughly 1~MK. In addition, a bidirectional jet is formed, as plasma is
rapidly accelerated away from the reconnecting region.

Since the lines of opposing polarity are located quite close to each
other around the null point, only a small disturbance is required
to start off the reconnection process the first time. This
happens soon after the start of the simulation, as fast mode
waves generated by the piston in
the central flux tube propagate upwards and reach the transition
region in as little as 10-12~seconds. There, they are refracted away from
the high Alfv\'{e}n speed region in the centre of the box (see
Fig.~\ref{firstwave})
and move instead towards the sides, where they reach the magnetic
null points and push the field lines together, triggering reconnection.

The piston, driving generally
along the field in a magnetically dominated (i.e. low-$\beta$) plasma,
is more efficient at generating slow mode than fast mode waves, so
these initial fast disturbances have low amplitude --- around 1~km~s$^{-1}$
just before reaching the transition region --- but this is enough to push
the field lines above and below the null point into each other and
trigger a weak reconnection event. This leads to moderate heating of the
reconnecting region and the formation of a bidirectional jet, with high
velocities upwards to the left and downwards to the right. The hot jet
going left then pushes cooler material sideways and upwards, propelling a
small, spicule-like protrusion of cooler chromospheric material into
the corona (Fig.~\ref{rjsecs}, {\it left panel}).

After some 160~seconds, or half a driver period later, the piston driven
disturbance
changes sign, and starts pulling the field lines above and below the null
point away from each other. Instead, the lines to the left and right of
the null point are pushed together, and reconnection occurs again, this
time in a direction perpendicular to the original reconnection event.
Another bidirectional jet is formed, this time being mostly vertical, and
the region around the null point (now embedded entirely in cooler
chromospheric gas) is heated to coronal temperatures. This heated
region spreads out along the jets and the magnetic field, and creates
a butterfly-shaped region of heated plasma (Figure~\ref{rjsecs}, {\it left
centre panel}).

The orientation of the upwards-moving jet changes as the reconnection
continues, and it rotates from being directed upwards to the right
(mainly across the field lines) to being directed somewhat to the left
(mainly along the field lines). The latter orientation allows the gas
to move much more freely, and the hot plasma is soon accelerated upwards
along the field lines with great speed (Fig.~\ref{rjsecs}, {\it right center
panel}). As it passes the now mainly
sideways moving cool jet generated by the first reconnection, it pushes
some of that cool gas upwards as well and greatly extends the length
of the cool jet, which reaches a maximum height of $z=8$~Mm.

This pattern then repeats itself, with reconnection being triggered every
150~s (i.e. every half period), alternating between vertical and horizontal
orientation, and propelling both cool and hot jets upwards. The later
events are not as strong as the first one, and are quite similar to
each other. An example of a later horizontally oriented reconnection jet
is shown in the right panel of Figure~\ref{rjsecs}.

\pagebreak

\subsubsection{Synthesised limb observations}

\begin{figure}
 \begin{center}
  \plotone{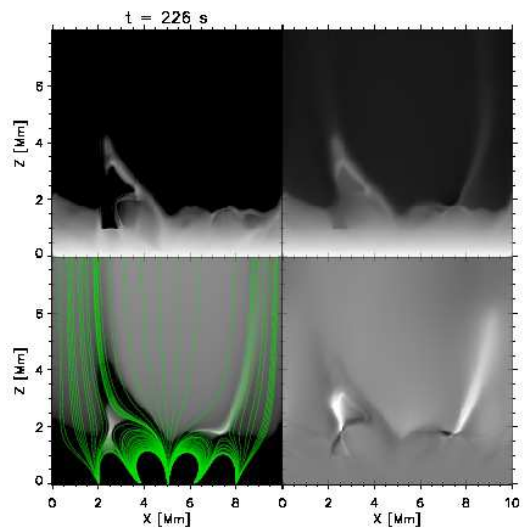}
 \end{center}
 \caption{Plots of the Ca signal ({\it top left}), density ({\it top right}),
          temperature
          with magnetic field lines ({\it bottom left}), and field-aligned
          velocity ({\it bottom right}) in the case with transition region
          null points and piston driver. The elapsed time is 226~s.
          Reconnection is producing a butterfly-shaped, vertically oriented
          heated region and jet on the left, and a hot TR jet on the right.
          This figure is available as a movie in the electronic version of
          the Journal.}
 \label{rjquad1}
\end{figure}

\begin{figure}
 \begin{center}
  \plotone{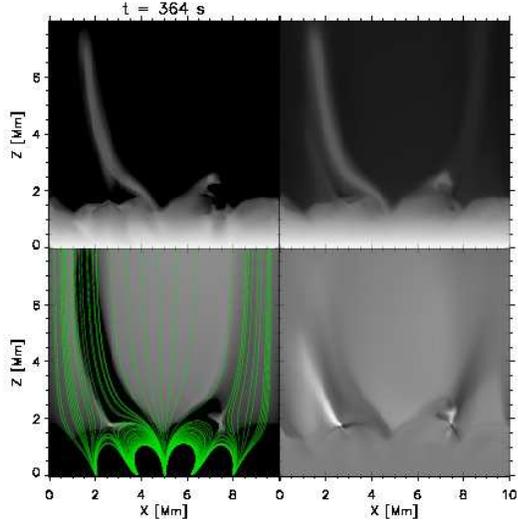}
 \end{center}
 \caption{Plots of the Ca signal ({\it top left}), density ({\it top right}),
          temperature
          with magnetic field lines ({\it bottom left}), and field-aligned
          velocity ({\it bottom right}) in the case with transition region
          null points and piston driver. The elapsed time is 364~s. The
          spicule-like cool jet on the left has reached its greatest length,
          extending nearly 8~Mm above the photosphere.}
 \label{rjquad2}
\end{figure}

\begin{figure}
 \begin{center}
  \plotone{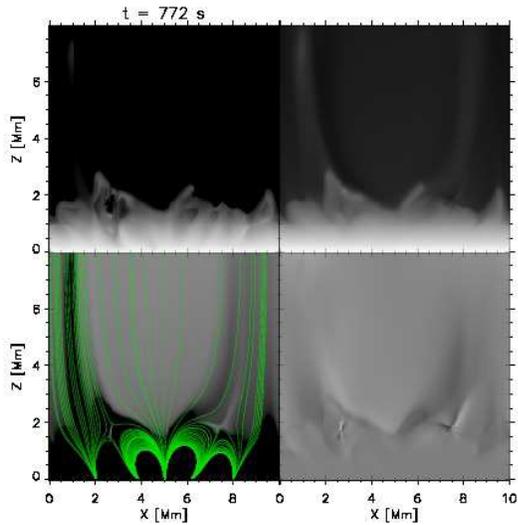}
 \end{center}
 \caption{Plots of the Ca signal ({\it top left}), density ({\it top right}),
          temperature
          with magnetic field lines ({\it bottom left}), and field-aligned
          velocity ({\it bottom right}) in the case with transition region
          null points and piston driver. The elapsed time is 772~s. A thin,
          low-density cool jet can be seen on the left in the temperature
          plot.}
 \label{thinjet}
\end{figure}

In order to get an impression of how these jets would look if observed
at the solar limb, we have used the non-LTE radiative transfer code
MULTI \citep{Carlsson1986} to calculate the population densities of
the different excitation levels of Ca~{\sc ii} ions in our simulations.
In the side views, we have chosen to concentrate on the Ca~{\sc ii}~H
line at 3968~\AA. In a 2D model, it is not possible to
do a full radiative transfer calculation along a line perpendicular
to the plane of the computational box, but the population density
of the upper level of this transition serves as a rough
approximation of the actual intensity.

In Figure~\ref{rjquad1}, we show plots of the logarithm of the calculated 
Ca~{\sc ii}~H upper level density ({\it upper left}, henceforth and in the
figure captions referred to as the Ca signal), the logarithm of the mass
density from the MHD simulation data ({\it upper right}), the plasma
temperature with overplotted
magnetic field lines ({\it lower left}), and the velocity along the magnetic
field ({\it lower right}) after 226~s. At this time, reconnection with
vertically oriented jets
is happening near the left-hand null point, while a horizontally oriented
event is happening near the right-hand null point --- though the jet on
the right-hand side of it is field-aligned and has turned upwards quite
sharply. The cooler jets show up in Ca~{\sc ii} as rather thin, elongated
features, similar to spicules or fibrils. In contrast, the hotter jets,
and the reconnection regions themselves, show no Ca~{\sc ii} signal.  The hot
jet on the right does have a density enhancement ({\it upper right panel}), but
it and the regions around the null points
are heated to temperatures of about 1~MK, far above the temperature where
calcium gets multiply ionised, which explains the lack of Ca~{\sc ii} emission.
The high velocities of the reconnection
jets are clearly visible in the lower right panel.

The cooler jets are seen to rise and fall; in Figure~\ref{rjquad2}, we show
the situation after 364~s, when the cool jet seen on the left in
Figure~\ref{rjquad1} has reached its maximum extent, after being pushed
upwards by the hot jet from the reconnection event at 200-240~s (ref.
Fig.~\ref{rjsecs}). It
reaches a height of nearly 8~Mm.

In Figure~\ref{thinjet}, we see a later, thinner cool jet on the left,
which shows up only weakly in Ca~{\sc ii}. The density plot ({\it upper right})
tells us why: its density appears to be too low, even though it is in about
the right temperature range.

The typical lifetimes of the cool jets are 200-300~s, and they reach maximum
lengths of 6-8~Mm above the photosphere.
Their thickness varies a lot with time, but
is typically slightly less than 1~Mm. These figures are roughly within
the ranges reported for spicules \citep{Beckers1968}. However, it is
important to look at the whole range of
spectral, spatial and temporal data available before trying to establish
correspondence between simulations and observations.

\subsubsection{Synthesised disk observations}

In producing synthesised disk observations, we look into the simulation
box directly from above, in the plane of the box. A proper radiative transfer
treatment is then possible, and has been carried out for the Ca~{\sc ii}~IR
line at 8542~\AA, widely used in spectroscopic studies. The calculations
have been made column by column, i.e. neglecting any radiative interaction
in the $x$-direction. A simplified
treatment based only on collisional excitation has been carried out
for the optically thin lines of C~{\sc iv} (1548~\AA), O~{\sc vi} (1032~\AA),
and Fe~{\sc xii} (195~\AA). These latter lines will typically be formed in 
the transition region or lower corona, whereas the Ca~IR line is formed in
the chromosphere. In the treatment of the Fe line, which has a wavelength
shorter than the Lyman threshold at 912~\AA, we have included the
effects of absorption by neutral hydrogen and neutral and singly ionised
helium.

\begin{figure}
 \begin{center}
  \plotone{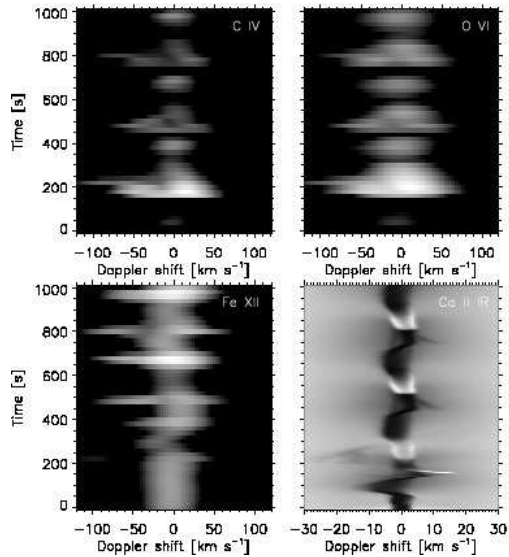}
 \end{center}
 \caption{Spectra of the high null point piston driver case, centered at
          $x=2.6$~Mm and with blueshifts
          shown as negative velocity. The C, O and Fe lines use logarithmic
          scaling in order to bring out the details of the much weaker later
          events. The difference between white (maximum intensity) and black
          (intensity cutoff) is a factor $10^4$. The UV lines experience
          strong brightening and extensive Doppler shifts when the reconnection
          events happen. The profiles are asymmetric, having higher blueshifts
          than redshifts, but the intensity maximum is slightly redshifted.
          The Ca spectrum uses linear scaling and a much smaller Doppler shift
          scale.}
 \label{rjspecs}
\end{figure}

$\lambda$-t plots of these lines, looking directly down on the centre of the
reconnection region at $x=2.6$~Mm, are shown in Figure~\ref{rjspecs}. The C,
O, and Fe lines have had their resolution and cadence downgraded to typical
instrumental values of 650~km (about 1'') and 20~s, and are plotted using
logarithmic scaling to bring out details in the weaker later events. Ca
is plotted at 150~km resolution and 2~s cadence, using linear scaling.

In the transition region lines, there is a sudden onset of
a very bright, very broad feature, with upflows of 80-90~km~s$^{-1}$ and
downflows of 60-80~km~s$^{-1}$. This happens first in C~{\sc iv}, the
lowest temperature line, at 160~s, then in O~{\sc vi} at 180~s (with a weaker
signal at 160~s).
This feature is caused by the sudden heating and the powerful
bidirectional jet generated by the reconnection. These spectra show
strong similarities to observations of explosive events
\citep{Innes+etal1997,Chae+etal2000}, which are believed to be caused 
by reconnection.
\citet{Dere1994} finds that explosive events have typical lifetimes
of 60~s, spatial extents of about 1.5~Mm, and typical Doppler shifts of
100~km~s$^{-1}$, though both higher and lower durations and Doppler shifts
have been reported. The figures match up well with our results; the main
parts of the brightenings fade away after 60~s and the spatial extent is
roughly in the quoted range (see Figs.~\ref{rjsecs} and \ref{rjquad1}).

The profiles are asymmetric, with larger blueshifts than redshifts, and
in particular the C and O lines have a short-lived blueshift of
140-150~km~s$^{-1}$ at 220~s. On the other hand, the intensity maximum
is redshifted. This behaviour can be readily understood in terms of the
density stratification. The reconnection outflow jets are expected to
propagate outwards at the Alfv\'{e}n speed, which is inversely proportional
to the square root of the density. Hence, the jets propagating upwards in
a stratified medium should be faster than those propagating downwards,
leading to an asymmetry like the one we find. The blueshift maximum
between 220 and 240~s happens when the heated jet, which is initially
embedded in material at chromospheric temperatures (Fig.~\ref{rjquad1}),
enters the corona where the Alfv\'{e}n speed increases markedly.
Meanwhile, the compression effect of the jet should be stronger for the
part propagating downwards into denser material, leading to higher
emission at slight redshifts.

The temperature in the reconnection region gets high enough to produce
significant emission also in the Fe~{\sc xii} line, but it is quite
strongly affected by neutral hydrogen absorption. The first powerful
event is completely absorbed because the reconnection region at
that time is located under a small jet with at least 1~Mm of chromospheric
material above it. The first horizontal jet reconnection event shows up
weakly at around 400~s, but strong blue- and redshifts are not seen
until the vertical jet at 500~s. The intensity maximum is in a horizontal
jet at around 680~s, at which time there is hardly any chromospheric
material above the reconnection region.

The chromospheric Ca~{\sc ii}~IR line shows signs of activity before
any of the higher temperature lines. A relatively strong upflow
(15~km~s$^{-1}$)
is caused after less than 100~s by the first weak reconnection episode.
This upflow is soon reversed, and after 150~s the first major reconnection
happens, changing the line centre from a redshift of 20~km~s$^{-1}$ to 0
almost instantly, while also yielding a significant blueshifted component
with speeds up to 30~km~s$^{-1}$.

After this violent first event, several periodic weaker reconnection events
follow. In these, the line only undergoes significant redshifts, with very
little evidence of blueshifts. This indicates that the line's main formation
height is below the reconnection region. The other lines show periodic
weaker brightenings, though still with quite high velocity. In every other
event, the main jets are oriented horizontally, leading
to smaller Doppler shifts. The intensity decrease is mainly due to reduced
density in the reconnection region; the reconnection jets cause a large
outflow of material, and because the inflow (amplified by the waves) soon
triggers another reconnection event, the area never recovers its initial
density.

\subsection{Chromospheric null points, piston driver}
\label{lpcase}

\subsubsection{General description}

These simulations use a piston driver of the same period (300~s) and
amplitude (1.1~km~s$^{-1}$) as in the previous section. However, here
the magnetic
null points are located around 450~km lower in the atmosphere, putting
them in the upper chromosphere rather than in the lower transition region.
The basic geometry of the magnetic field is not changed, as we can see
in Figure~\ref{mzprecon}; the only difference is a translation of the whole
geometry to lower heights.

\subsubsection{Synthesised limb observations}

\begin{figure}
 \begin{center}
  \plotone{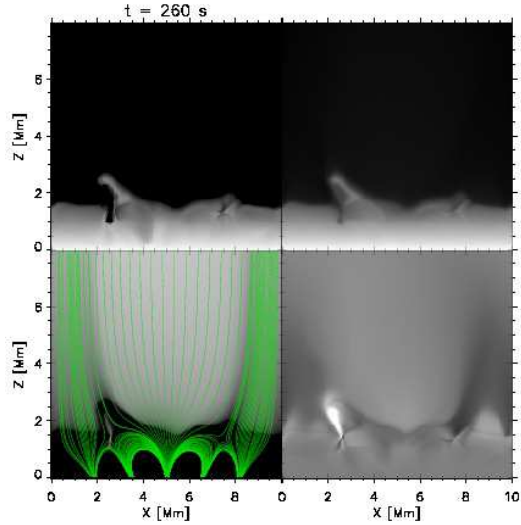}
 \end{center}
 \caption{Plots of the Ca signal ({\it top left}), density ({\it top right}),
          temperature
          with magnetic field lines ({\it bottom left}), and field-aligned
          velocity ({\it bottom right}) in the case with chromospheric null
          points and piston driver. The elapsed time is 260~s. The outflow
          speed from the reconnection region on the left is at its maximum.
          This figure is available as a movie in the electronic version of
          the Journal.}
 \label{mzprecon}
\end{figure}

\begin{figure}
 \begin{center}
  \plotone{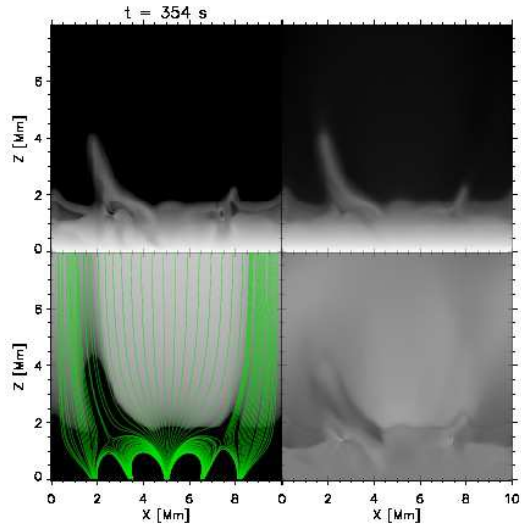}
 \end{center}
 \caption{Plots of the Ca signal ({\it top left}), density ({\it top right}),
          temperature
          with magnetic field lines ({\it bottom left}), and field-aligned
          velocity ({\it bottom right}) in the case with chromospheric null
          points and piston driver. The elapsed time is 354~s. The extent
          of the cool jet is at its maximum, 4~Mm above the photosphere.}
 \label{mzpjet}
\end{figure}

Figure~\ref{mzprecon} shows the simulation, viewed from the side, at the time
when the first and most powerful reconnection event has reached its maximum,
at around 260~s elapsed time. Once more, we have plotted the Ca~H signal
({\it upper left}), the logarithm of the mass density ({\it upper right}),
the temperature ({\it lower left}) and the field-aligned velocity ({\it lower
right}).
We clearly see the butterfly-shaped region of higher temperature ({\it lower
left}) and the fast bidirectional jet ({\it lower right}), although the
latter is now going upwards much faster than downwards due to the plasma
being able to more easily propagate along the magnetic field. The beginnings
of a jet can be seen in Ca~H ({\it upper left}), although the heated region
is too hot (and to some extent too evacuated) to show any Ca signal.

In Figure~\ref{mzpjet}, we show the situation as the cool jet reaches its
maximum extent, 90~seconds later. As could be expected, the velocity
signal is weak at this point, and the region around the null point is
no longer heated to coronal temperatures. The jet shows up clearly in
Ca, but reaches a maximum height of only
about 4.5~Mm, as compared to 8~Mm when the null points are
located in the transition region.
The general pattern described above repeats itself periodically every
300~s, with progressively slower and shorter jets.

Overall, this case is quite similar to the case with higher null points
described in the previous section.
The jets are slower and shorter because of the higher density surrounding
the null points, resulting in lower acceleration for similar force.

\subsubsection{Synthesised disk observations}

\begin{figure}
 \begin{center}
  \plotone{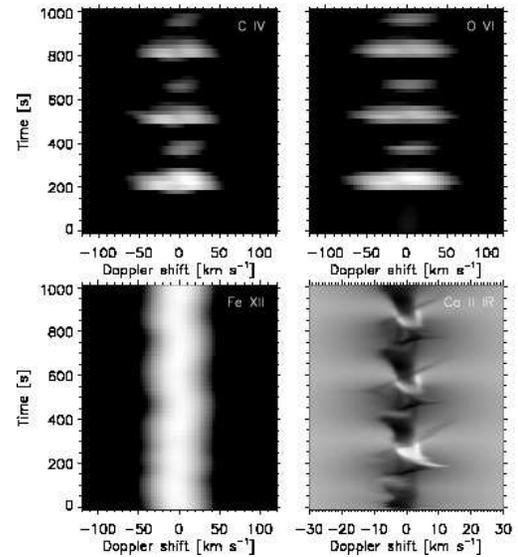}
 \end{center}
 \caption{Spectra of the chromospheric null point piston driver case,
          centered at $x=2.6$~Mm and using logarithmic
          scaling for the UV lines and linear scaling for Ca. C and O show
          largely symmetric profiles with maxima on each reconnection event
          with vertical jets. Fe shows only the background emission.
          Ca has a complex line profile with multiple components.}
 \label{mzspectra}
\end{figure}

\begin{figure}
 \begin{center}
  \plotone{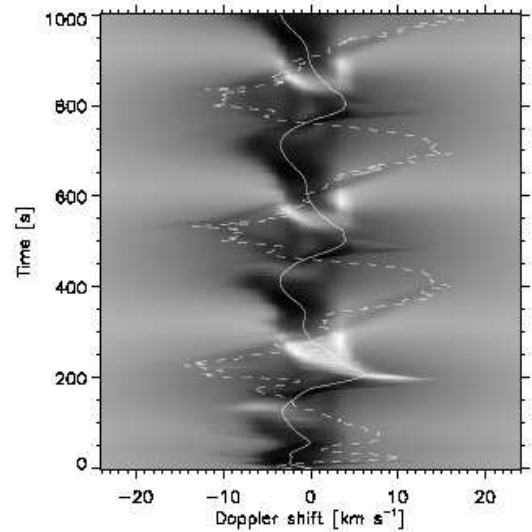}
 \end{center}
 \caption{Ca spectrum with doppler shifts ({\it greyscale}) in the
          chromospheric null
          point piston driver case, with velocities at $z=0.93$~Mm ({\it solid
          line}) and $z=1.37$~Mm ({\it dashed line}) superimposed. The former
          fits well with the line centre position, while the latter
          is a reasonable match for the blueshifted spikes and their
          following redshifts.}
 \label{mzcavel}
\end{figure}

Figure~\ref{mzspectra} shows $\lambda$-t diagrams of C~{\sc iv}, O~{\sc vi},
Fe~{\sc xii}
and Ca~{\sc ii}~IR as they would appear looking straight down at the centre
of the reconnection region ($x=2.6$~Mm), as for the high null point case
in section~\ref{hpcase}. The resolution is also the same as used previously.
In C~{\sc iv} and O~{\sc vi}, we get bidirectional
signals spaced 300~s apart, with Doppler shifts of 40-60~km~s$^{-1}$, but
little other signal. In Fe~{\sc xii}, the events give no visible signal,
and we just 
see the background emission. The reason for the low Fe emission is a
combination of the relatively lower temperature reached in the region
heated by reconnection and greater absorption by neutral hydrogen due
to the deeper location of the reconnection region. The UV lines are in
general more symmetric in this simulation,
because the hot jet does not reach the perturbed transition region and
corona before being cooled.

The Ca line, though, appears highly complex in this simulation. It
exhibits both absorption and emission, as well as strong Doppler shifts
into both red and blue. In observations, such a spectrum would be difficult
to make sense of. However, we have the full simulation data to work with,
including the velocity at all heights. In fact, at least two different
velocity components can be identified as being responsible for the
appearance of the spectrum, as shown in Figure~\ref{mzcavel}. The line
centre closely follows the vertical velocity at a height of 0.92~Mm,
below the magnetic null point, as marked with a solid white line in the
figure. The strongest blueshifted signal, which later passes through the
line centre and becomes redshifted, corresponds reasonably well to the vertical
velocity at a height of 1.37~Mm ({\it dashed line}), just above the null point,
which is located at 1.30~Mm.
These two velocity components are in counterphase as a result of the
bidirectional jet produced by the reconnection.

This nicely illustrates the fact that spectral lines are formed over a
range of heights, rather than at one specific height, and that care
must be taken when using them as probes of atmospheric conditions.
But it also illustrates that such a multi-component signal could be
used as evidence of reconnection via the characteristic bidirectional
velocity pattern, as long as the reconnection happens in the height range
where the line is formed. By contrast, in our simulation with null
points in the transition region, we observe only the downflows because
the line is primarily formed below the null points and reconnection region.

\subsection{Transition region null points, Hinode driver}
\label{hhcase}

\subsubsection{General description}

In this simulation, we go back to the magnetic configuration of the
first simulation studied (section~\ref{hpcase}), where the null points
are located in the
transition region. However, instead of using a localised, monochromatic
piston driver, we use velocity data obtained with Hinode/SP (as described
in the introduction) 
to drive the whole lower boundary. As before, we will not
attempt to match the simulation
to specific observations; the data simply serve as a more random and
realistic velocity driver with a wider range of frequencies than our
monochromatic piston.

\subsubsection{Synthesised limb observations}

\begin{figure}
 \begin{center}
  \plotone{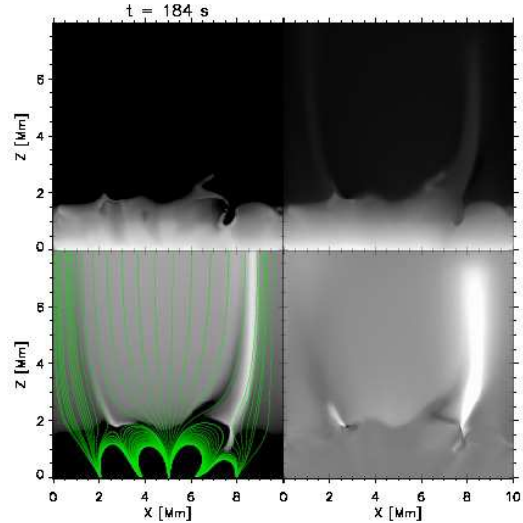}
 \end{center}
 \caption{Plots of the Ca signal ({\it top left}), density ({\it top right}),
          temperature
          with magnetic field lines ({\it bottom left}), and field-aligned
          velocity ({\it bottom right}) in the case with transition region
          null points and Hinode driver. The elapsed time is 184~s. A strong
          and very fast hot jet is being produced by the reconnection at the
          right-hand null point. This figure is available as a movie in the
          electronic version of the Journal.}
 \label{hinhotjet}
\end{figure}

\begin{figure}
 \begin{center}
  \plotone{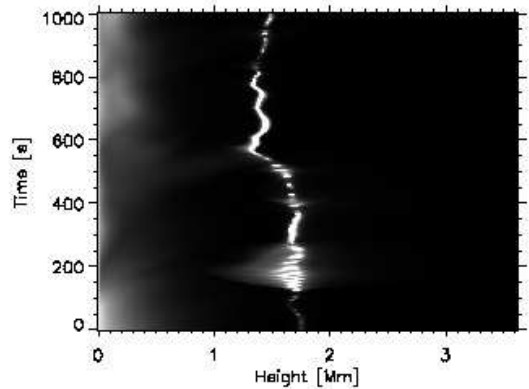}
 \end{center}
 \caption{$\beta$ plot showing the height change of the right-hand
          null point with time
          when it starts out in the transition region and the Hinode driver
          is applied.}
 \label{npheight}
\end{figure}

\begin{figure}
 \begin{center}
  \plotone{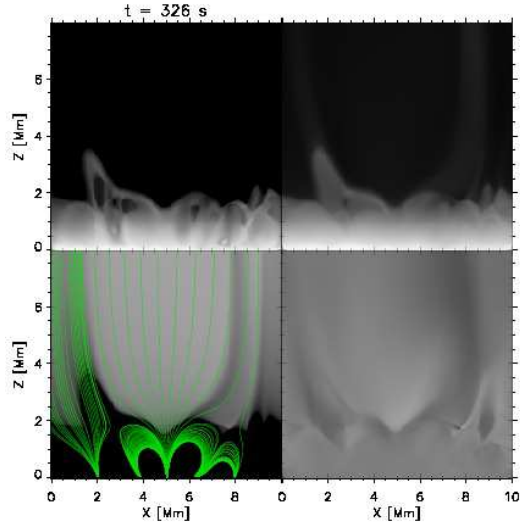}
 \end{center}
 \caption{Plots of the Ca signal ({\it top left}), density ({\it top right}),
          temperature
          with magnetic field lines ({\it bottom left}), and field-aligned
          velocity ({\it bottom right}) in the case with transition region
          null points and Hinode driver. The elapsed time is 326~s. This
          figure shows the longest of the cool jets that appear in this
          simulation.}
 \label{hincooljet}
\end{figure}

Using the Hinode driver changes a number of things. In general, the velocity
field in the atmosphere becomes much more complex. On the left-hand side,
the first reconnection
has its jet oriented vertically, as opposed to the horizontal orientation
we get in the runs with a piston driver. This is a result of the different
phase of the first waves to reach the null point. In this case, as the
reconnection happens in the transition region, only the hotter material above
the null point is ejected upwards, and there is little cool material above
the null point to
be accelerated by later reconnection events. As a result, only few and
short cool jets are formed at all.

At the right-hand side null point, no significant reconnection happens until
130~s, but then a quite major event occurs, with its jet oriented vertically.
This leads to the ejection of a powerful hot jet, as seen in
Figure~\ref{hinhotjet}. The jet is too hot to show up in Ca.

This latter reconnection event is actually quite long-lasting,
the flow direction only being reversed at 290~s, and then only for 90~s
before another 170~s period of vertically oriented reconnection. The
sustained event releases quite a bit of energy from the magnetic field and
changes its structure fairly significantly; among other things, the
null point is moved down into the chromosphere by 3-400~km
(Fig.~\ref{npheight}). Later jets are fairly weak, the cool ones
reaching only around
1~Mm above the transition region (Fig.~\ref{hincooljet}).

The long duration of the event is likely a result of the specifics of
the driver. On the right-hand side of the box, the input velocity only
rarely changes sign, leading to a fairly continuous flow in one direction,
rather than the periodic reversals of the piston driver.

\subsubsection{Synthesised disk observations}

\begin{figure}
 \begin{center}
  \plotone{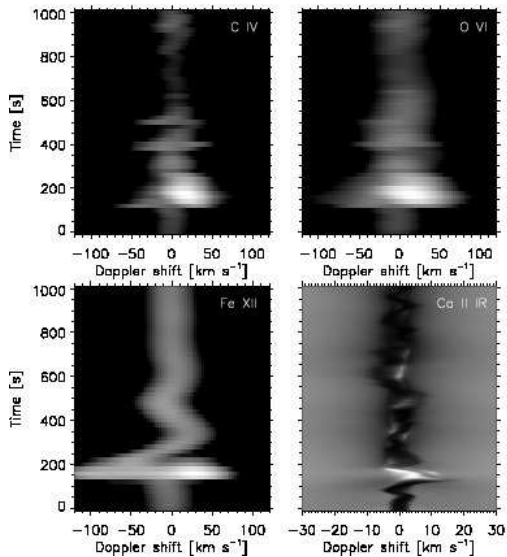}
 \end{center}
 \caption{Spectra of the high null point Hinode driver case, centered at
          $x=7.6$~Mm. The UV lines use logarithmic
          scaling and the Ca line uses linear scaling. The C and O
          lines show mainly redshifts because the matter moving upwards
          is heated out of their passbands. Fe shows not only the signal
          of the heated area around the reconnection point, but also of the
          fast hot jet to the side of it, shown in more detail in
          Fig.~\ref{hin168spec}.}
 \label{hin152spec}
\end{figure}

\begin{figure}
 \begin{center}
  \plotone{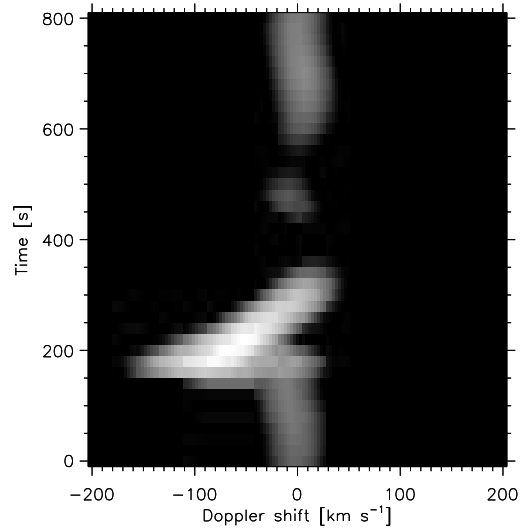}
 \end{center}
 \caption{Fe {\sc xii} spectrum of the high null point Hinode driver case,
          centered at $x=8.45$~Mm and using logarithmic scaling. This
          spectrum shows the very fast hot jet and its considerable
          blueshift of 160 km~s$^{-1}$.}
 \label{hin168spec}
\end{figure}

\begin{figure}
 \begin{center}
  \plotone{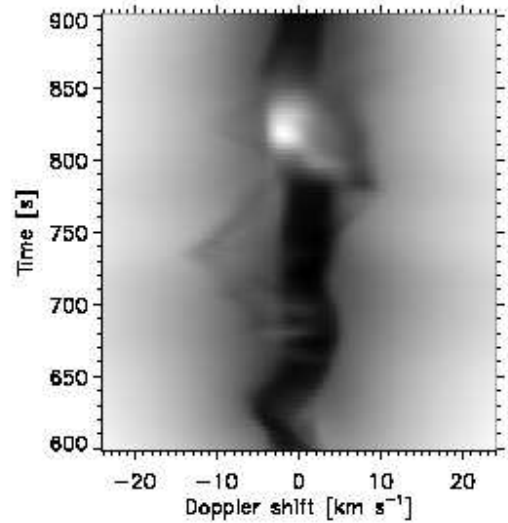}
 \end{center}
 \caption{Ca {\sc ii} IR spectrum of the high null point Hinode driver case,
          centered at $x=7.9$~Mm. Weak blueshifted excursions similar
          to those in Fig.~\ref{mzspectra} can be seen as a result
          of the null point having moved down after extended reconnection.}
 \label{hin158spec}
\end{figure}

Because the null point in this simulation displays significant movement,
both horizontally and vertically, and because the jets tend to follow the
magnetic field, which is slightly inclined even in the lower corona, we
can find interesting phenomena at several different locations when
looking down as we would do if the events happened on the solar disk.
We calculate the same C, O, Fe, and Ca lines as above, and use the
same spatial resolution and exposure times.

In Figure~\ref{hin152spec}, we show $\lambda$-t plots of the different
spectra centered at $x=7.6$~Mm. This is the place where the transition region
and coronal lines have their highest intensities, looking straight down
on the reconnection region. The C~{\sc iv} line shows little bidirectionality
in this case, having its maximum at redshifts of around 25~km~s$^{-1}$.
The O~{\sc vi} line is bidirectional, but has much higher intensity in the red.
The Fe~{\sc xii} line is more extended, but also has its maximum at the same
redshift.
This line not only includes emission from the region immediately surrounding
the null point, but also from the upwards propagating fast hot jet
between $x=8$ and $x=9$~Mm. A spectrum centered at $x=8.45$~Mm showing this
jet in more detail is shown in Figure~\ref{hin168spec}.
We can see the strong signature of the fast hot jet and its deceleration
between 160 and 340~s. It reaches an impressive blueshift of roughly
160~km~s$^{-1}$, matching the maximum field-aligned velocity from the
simulation data.

The Ca~{\sc ii}~IR line shows only redshifts for this event (130-200~s),
some of them
rather powerful (up to 20~km~s$^{-1}$), and exhibits strong
central reversal at around 140~s, when the plasma is significantly heated
by the reconnection. The later Ca spectrum is more complex,
likely showing several overlapping velocity components as in the low
null point piston driver case in section~\ref{lpcase}, as well as the
effects of the movement
of the null point itself. Of course, the velocity input from the Hinode
driver includes a wide spectrum of frequencies rather than the single
one of the piston, which serves to further complicate the simulated Ca
spectrum.

The effects of having the null point move several hundred km downwards
are visible in the Ca signal in some locations. In Figure~\ref{hin158spec},
we show a portion of the spectrum at $x=7.9$~Mm. Between 680 and 780~s, there
is a clear blueshifted excursion, similar to what we saw in the spectra
of the simulations with chromospheric null points (Figs.~\ref{mzspectra}
and \ref{mzcavel}). By the time of this event, the right-hand null point
in this simulation is at a height of $1.4$~Mm (Fig.~\ref{npheight}),
very close to the height of the null points in the other simulation.

\subsection{Chromospheric null points, Hinode driver}
\label{lhcase}

\subsubsection{General description}

\begin{figure}
 \begin{center}
  \plotone{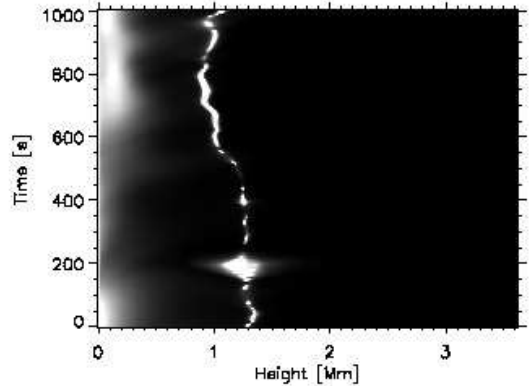}
 \end{center}
 \caption{$\beta$ plot showing the height change of the right-hand
          null point with time
          when it starts out in the chromosphere and the Hinode driver is
          applied.}
 \label{npheight2}
\end{figure}

In this simulation, we again use the magnetic configuration that
puts the magnetic null points in the chromosphere, but use the same Hinode
driver as in the previous section. Once again this leads to extensive
reconnection near
the right-hand null point, but because that is now embedded in chromospheric
plasma, it produces a major cool jet rather than the hot one produced when the
null points are in the transition region. Again the reconnection changes
the field and pushes the null point lower in the atmosphere after about
500~s (Fig.~\ref{npheight2}). Later jets are fairly weak and do not extend
far above the transition region; neither do the jets produced by the
weaker events at the left null point. Some jets are also produced by shock
waves propagating from the lower boundary, especially in the magnetic
channel connecting to $x=2$~Mm --- the shock fronts are clearly visible in
movies of the field-aligned velocity. This process is similar to the one
that produces dynamic fibrils \citep{Hansteen+etal2006}, but the jets
produced this way tend to reach heights of no more than $0.5$~Mm above
the transition region in this simulation. 

\subsubsection{Synthesised limb observations}

\begin{figure}
 \begin{center}
  \plotone{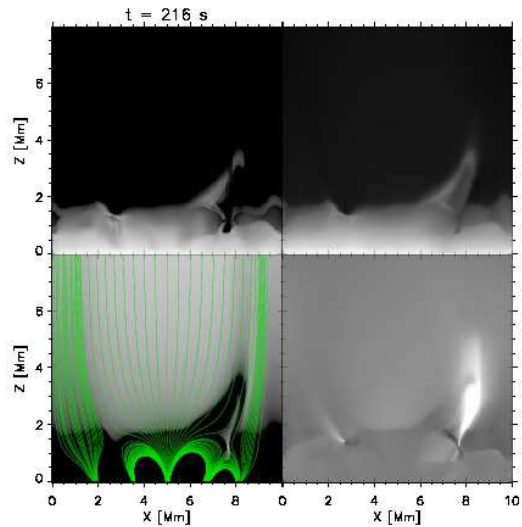}
 \end{center}
 \caption{Plots of the Ca signal ({\it top left}), density ({\it top right}),
          temperature
          with magnetic field lines ({\it bottom left}), and field-aligned
          velocity ({\it bottom right}) in the case with chromospheric
          null points and Hinode driver. The elapsed time is 216~s. A powerful
          reconnection event is happening at the right-hand null point,
          heating a large area and propelling a cool jet upwards at great
          speed. This figure is available as a movie in the electronic
          version of the Journal.}
 \label{hinzprecon}
\end{figure}

\begin{figure}
 \begin{center}
  \plotone{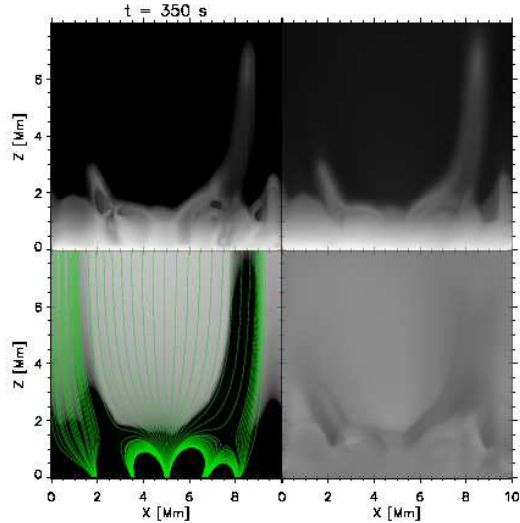}
 \end{center}
 \caption{Plots of the Ca signal ({\it top left}), density ({\it top right}),
          temperature
          with magnetic field lines ({\it bottom left}), and field-aligned
          velocity ({\it bottom right}) in the case with chromospheric
          null points and Hinode driver. The elapsed time is 350~s. The
          cool jet has reached its maximum extent, 8~Mm above the photosphere.
          \vspace{2mm}}
 \label{hinzpjet}
\end{figure}

In Figure~\ref{hinzprecon}, we show the Ca~H signal, density, temperature
and field-aligned velocity at 216~s elapsed time, when the region heated
by the first powerful reconnection event has reached its maximum extent.
We see that the upper part of that region follows the field lines upwards;
the total length of the heated region is more than 2~Mm. We also see the
high velocity of the jet. Note that the heated region gets too hot to
show up in the Ca signal.

In Figure~\ref{hinzpjet}, we show the situation at 350~s, when the cool
jet has reached its maximum extent. As we see, it reaches a height of nearly
8~Mm above the lower boundary, or 6~Mm above the transition region.
This cool jet shows up quite strongly in Ca, and has a noticeable density
enhancement compared to the surrounding corona. The velocity, as one would
expect, is close to zero at this time.

This jet does not form a perfect parabola in a $z$-$t$ diagram
(it ascends slightly faster than it descends), but the best parabolic fit,
using the method of \citet{DePontieu+etal2007} and \citet{Heggland+etal2007},
yields a maximum velocity $v_{max}$ of 52~km~s$^{-1}$, a deceleration
$d$ of 250~m~s$^{-2}$, and a duration $P$ of roughly 430~s. Although
these values fall outside the range studied by \citet{Heggland+etal2007}
in their study of shock wave-driven dynamic fibrils --- in particular,
the maximum velocity is much higher than would reasonably develop
through steepening of acoustic waves alone --- they are
a pretty good fit to the formula predicted and found in that paper,
namely \begin{equation} d = \frac{v_{max}}{P/2}; \end{equation} not too
surprisingly since the jet is driven by a shock, albeit one generated by
reconnection rather than directly by photospheric motions.

\subsubsection{Synthesised disk observations}

\begin{figure}
 \begin{center}
  \plotone{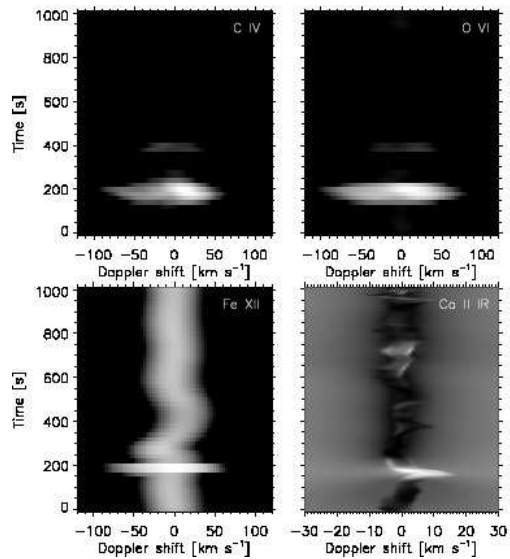}
 \end{center}
 \caption{Spectra of the chromospheric null point Hinode driver case, centered
          at $x=7.65$~Mm. The UV lines use logarithmic scaling and the Ca
          line linear scaling. All lines have huge maxima when the first
          major reconnection event happens. Fe is largely symmetric
          while the C and O lines have asymmetries similar to those in
          Fig.~\ref{rjspecs}. Ca, formed at the lowest height, is
          redshifted, but the later signal is shifted more towards
          the blue after the null point moves to lower heights.}
 \label{hmzspectra}
\end{figure}

In Figure~\ref{hmzspectra}, we show the simulated C~{\sc iv}, O~{\sc vi},
Fe~{\sc xii}, and Ca~{\sc ii}~IR spectra
looking down on $x=7.65$~Mm. As in the other cases, the TR lines are plotted
at 650~km and 20~s resolution, and the Ca spectrum at 150~km and 2~s.

C, O, and Fe all have huge maxima in connection with the first reconnection
event, centered at 180-200~s --- even with a lower intensity cutoff of
$10^{-4}$ times the maximum intensity, as in all the UV spectral plots,
only Fe shows any part of the background signal.
The event shows up as a clear bidirectional jet in all three, with
50-70~km~s$^{-1}$ velocities; even up to 90~km~s$^{-1}$ in O~{\sc vi}.
C and O have some of the same asymmetry as in the TR null point piston
driver case (section~\ref{hpcase}), with slightly higher blueshifts
than redshifts and slightly redshifted intensity maxima.

In Ca, we get strong emission and a redshift of 17~km~s$^{-1}$ for this
event. Later on, the spectrum gets more complex, as usual with
the Hinode driver. However, it is notable that after the null point moves
down at around 500~s, the Ca spectrum mainly shows blueshifts, since
the primary formation height is now above the reconnection region. This
makes a nice contrast with the mainly redshifted spectrum of the high
null point piston driver case (Fig.~\ref{rjspecs}), where the null point
is located almost 900~km higher up.

\subsection{Energy release estimates}

\begin{figure}
 \begin{center}
  \plotone{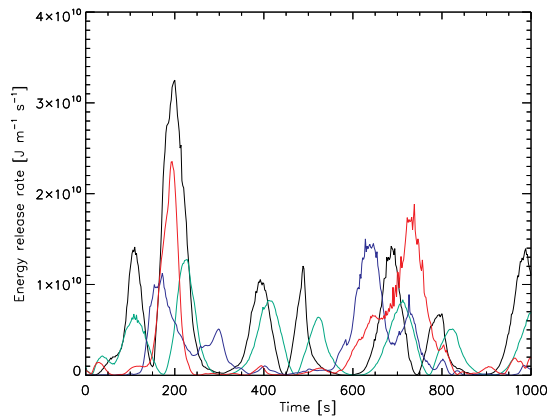}
 \end{center}
 \caption{Energy release rate in the form of joule heating in the four
          different cases: transition region null point with piston driver
          ({\it black}), chromospheric null point with piston driver
          ({\it green}),
          transition region null point with Hinode driver ({\it blue}), and
          chromospheric null point with Hinode driver ({\it red}).}
 \label{jouleheat}
\end{figure}

In Figure~\ref{jouleheat}, we have plotted the energy release rate in
the form of joule heating ($j^2/\sigma$, where $j$ is the current density
and $\sigma$ the conductivity) in our four simulations. The energy release
is calculated over the reconnection region, in this case defined as the
region around the null point where the joule heating is above 5\% of the
value at the null point. The results are not very sensitive to the particular
value of the threshold, and changing it from 1\% to 10\% changes the
calculated energy release by less than 10\%. In the piston driver cases,
we include the region around the left-hand null point, while we use the
right-hand one in the Hinode driver cases.

As we would expect from the side views and spectra, the greatest peaks in
the energy release occur in the transition region null point piston driver
case ({\it black line}) and the chromospheric null point Hinode driver case
({\it red line}). The two piston driver cases are more regular, with double
peaks (horizontal, then vertical jets) appearing every 300~s. The Hinode driver
cases have second peaks appearing between 600 and 800~s, and the energy
release is actually greater in these events than in the first one at
around 200~s, but the later events
show only weak observational signatures. This is because the jets of these
events are horizontally oriented, giving no Doppler shifts in the simulated
spectra and not very tall jets, and also because the energy release happens
over a longer period of time than at the first peak, at least in the case
with chromospheric null points.

The energy release rate is given in J~m$^{-1}$~s$^{-1}$ because the total
energy release will be dependent on the size of the reconnection region
in the third dimension. In order to calculate the total energy release,
we must assume an extent, and this value will necessarily be somewhat
arbitrary. A value of $0.5$~Mm has been assumed in Table~\ref{heattab}.
The released energy per event is comparable to the generally assumed
values for nanoflares, though not all the energy released in these
events will be in the form of joule heating.

\begin{deluxetable}{lrcrc}
\tablecolumns{5}
\tablewidth{0pt}
\tablecaption{Energy release per reconnection event \label{heattab}}
\tablehead{
  \colhead{} & \multicolumn{2}{c}{First event} & \multicolumn{2}{c}{Strongest later event} \\
  \colhead{Case} & \colhead{Energy [J]} & \colhead{Duration [s]} & \colhead{Energy [J]} & \colhead{Duration [s]}}
\startdata
TR np/piston & $1.23 \times 10^{18}$ & 280 & $5.40 \times 10^{17}$ & 280 \\
Ch$.$ np/piston & $5.48 \times 10^{17}$ & 280 & $4.19 \times 10^{17}$ & 280 \\
TR np/Hinode & $5.01 \times 10^{17}$ & 240 & $7.15 \times 10^{17}$ & 260 \\
Ch$.$ np/Hinode & $3.31 \times 10^{17}$ & 100 & $7.18 \times 10^{17}$ & 280
\enddata
\end{deluxetable}

\section{Discussion and summary}

These simulations have been able to produce a number of jet phenomena
through the same basic mechanism of wave-induced, or in the Hinode driver
cases also convective flow-induced, magnetic reconnection.
In particular, they appear to be the first 2D simulations to show the
formation of spicule-like jets as a result of reconnection, while including
most important physics; the main omission is time-dependent ionisation.
The simulated spicules match observed lifetimes and lengths.

In this model of spicule formation, we also expect a brightening and
broadening of UV lines at the footpoint of the spicule at the start of
its lifetime; the heated plasma causing this is in general cooled down
on a shorter timescale than the lifetime of the spicule. This UV bright
point is a property our model shares with most earlier models that
assume a sudden energy/pressure deposition as the source of spicules,
rather than a velocity perturbation at photospheric heights. Although
UV bright points are frequently observed in connection with spicules,
some observations \citep[e.g.][]{Suematsu+etal1995} indicate that this
brightening happens after the ascending phase of the spicule, rather
than at the beginning. This timing problem is not resolved by our model.

The UV brightenings that are produced in the reconnection regions
have many of the same properties as observed explosive events, which
could point towards a possible connection between spicules and explosive
events, as suggested by \citet{Wilhelm2000}. One of the plasma
acceleration mechanisms suggested in that paper, a slingshot effect
\citep[ Fig. 8]{Wilhelm2000} may be happening at times in our model;
see for example the left centre panel of Figure~\ref{rjsecs}, where
the mass flow is perpendicular to the field lines and chromospheric
matter is propelled upwards by several hundred km. On the other hand,
flow along the field proves much more effective in lifting chromospheric
material in our simulations (right centre panel of Fig.~\ref{rjsecs} and
lower two panels of Fig.~\ref{rjquad1}); different field geometries
could give different results.

Although our UV brightenings match many of the properties of explosive
events, there are also some differences. Our events are generally very
bright, with maximum intensities 50-100 times higher than the ``quiet''
profiles --- in some cases, even 1000 times. Although some explosive
events have strong brightenings, the majority are not particularly bright
--- \citet{Innes2001} quotes typical brightening factors of 2-5. Also,
we get significant signal in the lower coronal Fe~{\sc xii} 195~\AA{} line,
while explosive events rarely show up in coronal lines.
\citet{Erdelyi+etal1999} show one example of an explosive event brightening
in TRACE 171~\AA{}, but since it does not show up in the lower temperature
Mg~{\sc x} 625~\AA{} line, they interpret it as a brightening of a transition
region O~{\sc vi} line within the passband rather than the coronal Fe~{\sc x}
line.

In one of our cases (Fig.~\ref{mzspectra}), we also find only very weak
Fe~{\sc xii}
emission. Since the heating is quite strongly localised to the reconnection
region, it becomes a blob of coronal plasma surrounded by low-temperature
chromospheric material. In the low-temperature region, neutral hydrogen
is present in significant proportions, and this leads to strong Lyman
absorption of all lines with shorter wavelengths than 912~\AA{}. This,
combined with the heating events themselves being relatively weak, explains
the lack of Fe emission in this case, and could also do the same for
observed explosive events if in fact they occur relatively deep in the
chromosphere. However, the C and O lines will be unaffected by this
absorption, and are still stronger than observed. Also, the high velocities
observed in many explosive events indicate a quite high formation height.
A more likely explanation for the discrepancy may therefore be that our
reconnection events are simply more energetic than those responsible for
most explosive events.
The temperatures reached are high enough ($> 1$~MK) that they might even
have an X-ray signal.

It should be noted that even the very strong first event in the TR null point
piston driver case is completely absorbed by the neutral hydrogen above it.
Therefore, the absence of observed signal in high-temperature lines does not
necessarily mean that the plasma is not heated to such temperatures,
as the emission could be absorbed by overlying cool matter.

Finally, wave-induced reconnection offers a natural explanation for
the repetitive behaviour sometimes observed in explosive events
\citep[e.g.][]{Chae+etal1998a, Ning+etal2004}. If these events are
caused by reconnection induced by waves,
we would naturally expect repetition at time scales of 3-5~minutes,
corresponding to the dominant periods in the solar chromosphere and
photosphere.

Although we find a number of similarities between our simulation
results and observations,
the complexity of the signals in our synthetic spectra and
images shows that it is not surprising that the many different jet-like
phenomena that are observed in the solar atmosphere have been, and
remain to some extent, such a puzzle. For example, in the same
simulations, with similar magnetic field geometry and driver, we find
at different times jets that show significant chromospheric signal
without transition region or coronal counterparts, jets that show
only blueshifts in
the chromosphere, jets that only show redshifts in the chromosphere,
jets that do not show chromospheric signatures but are dominated by
bidirectional flows in the transition region, and jets in the
transition region lines that do not show
any coronal counterpart. This is a natural consequence of the
complex mix of magnetic field geometry, the history of the plasma motions,
the narrow height/temperature range in which the observables are
formed, and the varying mix of heating and acceleration in reconnection
events. As a result, establishing the relationship between
various types of events in the solar atmosphere is fraught with
difficulties that can only be resolved by statistical comparisons of
spatio-temporal data of high quality with advanced
radiative MHD models.

\acknowledgements

This work was supported by the Research Council of Norway through grants
159483/V30, 170926, and 170935, as well as a grant of computing time from
the Program for Supercomputing. B.~D.~P. was supported by NASA grants
NNG06GG79G, NNX08AL22G, and  NNM07AA01C (Hinode). L.~H. thanks \O ystein
Langangen and Mats Carlsson for useful discussions.


\end{document}